# A tour about Isaac Newton's life

**A.C. Sparavigna[1] and R. Marazzato[2]**
1 Department of Applied Science and Technology, Politecnico di Torino, Italy
2 Department of Control and Computer Engineering, Politecnico di Torino, Italy

*Here we propose a tour about the life of Isaac Newton, using a georeferenced method, based on the free satellite maps. Our tour is modelled on the time-line of the great scientist's life, as an ancient "itinerarium" was modelled on the Roman roads, providing a listing of places and intervening distances, sometimes with short description or symbols concerning the places. KML language and Google Earth, with its Street View and 3D images are powerful tools to create this virtual tour.*

**Keywords**: Georeferencing, Satellite Maps, KML, XML, Acme Mapper, History of Physics.

In a recent paper we have proposed to give georeferences for the lives of famous people [1]. We suggested to study in such a manner the lives of scientists for teaching purposes, to increase the appeal of scientific disciplines. That paper was just a proposal: here we are discussing an implementation about the Isaac Newton's life in KML language.
The best theoretical model which we can use to describe our aim and method is the ancient roman "itinerarium". The *itineraria* were quite simple maps in the form of a listing of cities, villages and even refreshment places, connected by lines and reporting the intervening distances [2]. In its original form, the itinerarium was simply a list of places along roads, with some specific information. As the source of all of them there was a stone-engraved one located near the Pantheon in Rome, containing enough information to create a suitable itinerary for the traveller. We used in the title of our manuscript the term "tour", because it is the modern term closer to the Latin term. In the case of a life, it is a time-line to rule the tour, not the locations.
The satellite images, which we can find on the World Wide Web in Google Earth, can become our modern map on which we mark the list of places and the related information of our tour about the life of a person. Here, it is the life of a great scientist, Isaac Newton. Our tour is not a biography; however, unlike a simple profile, it has the possibility to include detailed descriptions or other accounts, portraying the subject's experience at some specific locations. We can highlight the life from the point of view of the local geographic experience, following the places where this person lived or paid a visit. In practice, using some tags on a satellite map, we can find the places and then record and share, by listing them into a "placemarks" file. In the case that we are using the free version of Google Earth, the data can be saved in the KML format.
Keyhole Markup Language, KML, is the notation of a mark-up language (XML) for expressing geographic annotation and visualization within Internet-based maps [3]. This sort of geo-referencing a person's life is a new application of georeferencing methodologies, which are well known and used for the localization and evaluation of cultural heritage [3-6] and even for teaching physics [7-10].
Before starting our tour with Google Earth, let us note that some simple implementations can be made by means of ACME Mapper [11]. It is based on the same satellite imagery of Google Earth. ACME Mapper has the possibility to mark the places, and provides a list of all the marked places with their coordinates and distances. In Figure 1 for instance, the red markers are showing some places where Isaac Newton lived.

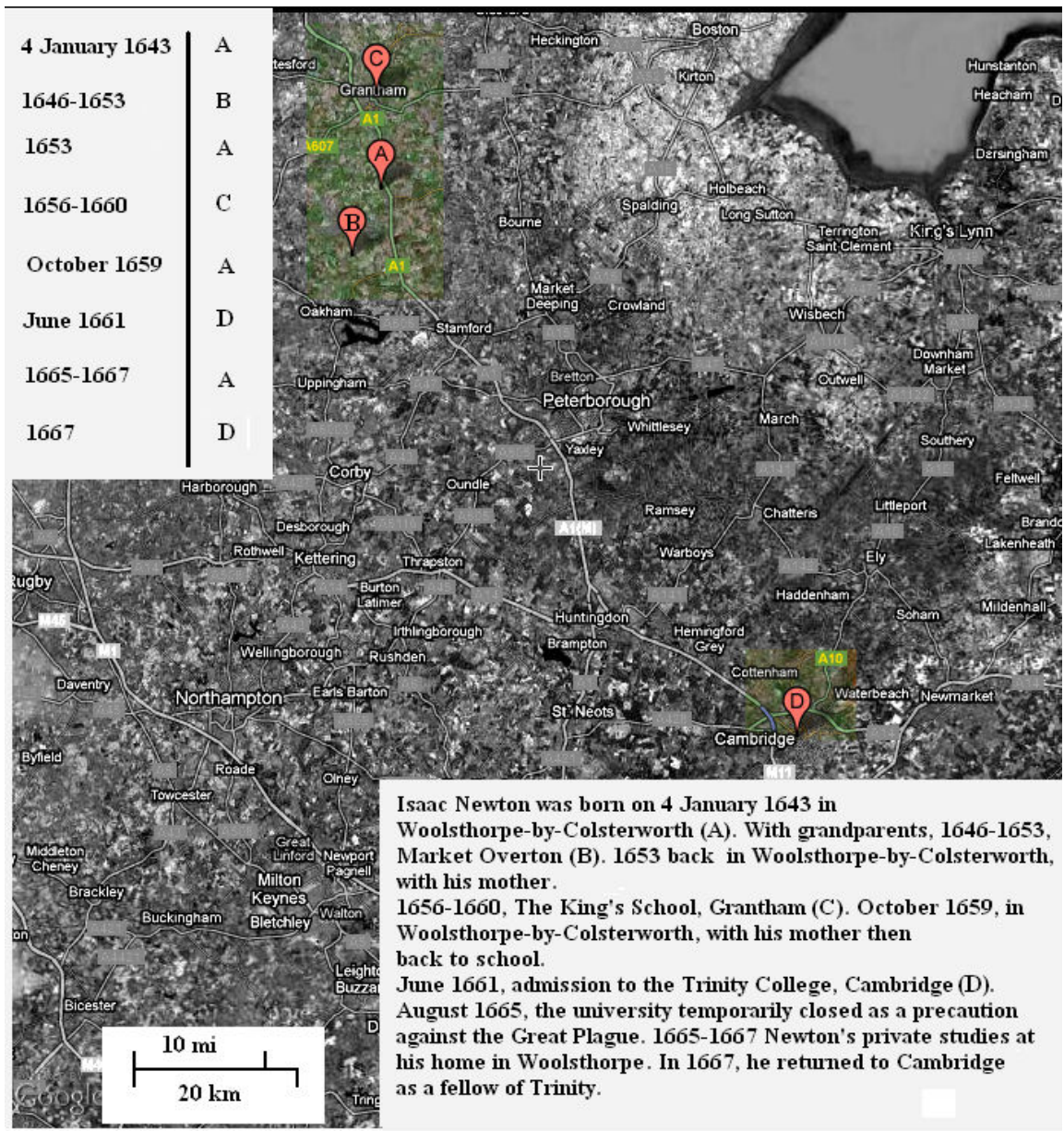

Fig.1: Acme Mapper of the first part of Newton's life. Up-left, there is the time-line, giving the periods of time that Newton spent in the locations A, B, C and D. Details on locations are given in the inset.

In Fig.1, on the left, there is the time-line of the first part of Newton's life. It is giving the periods of time that Isaac Newton spent at the locations shown on the map, locations that we have named A, B, C and D, during the first part of his life. Information about Newton is coming from Ref.12. Our itinerarium is therefore the time-line on the left, with details shown in the inset of the figure.
We see that he spent his youth in the country, near Nottingham, and that he studied at Cambridge. During the Great Plague he came back to Woolsthorpe, and there, in a garden he had the intuition of a universal gravitation observing a falling apple [12].
An approach using the Acme Mapper is rather simple, but it is impossible to use just one image to describe all the relevant locations. Moreover, the image is static, not allowing an interactive tour. The use of a KML on Google Earth is non-static, based on dynamic images, and therefore we can follow the time-line as a motion on the proper coordinates. We have therefore written a tour in KML, among the places related to the young Isaac.

The file can be downloaded from the site, Newton Tour, the reader can find at the following address http://staff.polito.it/roberto.marazzato/newton-tour/ (the first link of the site is to the latest version, http://staff.polito.it/roberto.marazzato/newton-tour/NewtonTour.kml).

Once the KML file is downloaded, it is possible to run it by means of Google Earth. After launching the file, a map appears showing the locations, listed on the left of the page. We can select a specific place from the list and pay a visit to. Choosing for instance Woolsthorpe Manor, the birthplace of Sir Isaac Newton is shown by a balloon, where there is a picture of it and a movie that one can watch on YouTube (Fig.2).

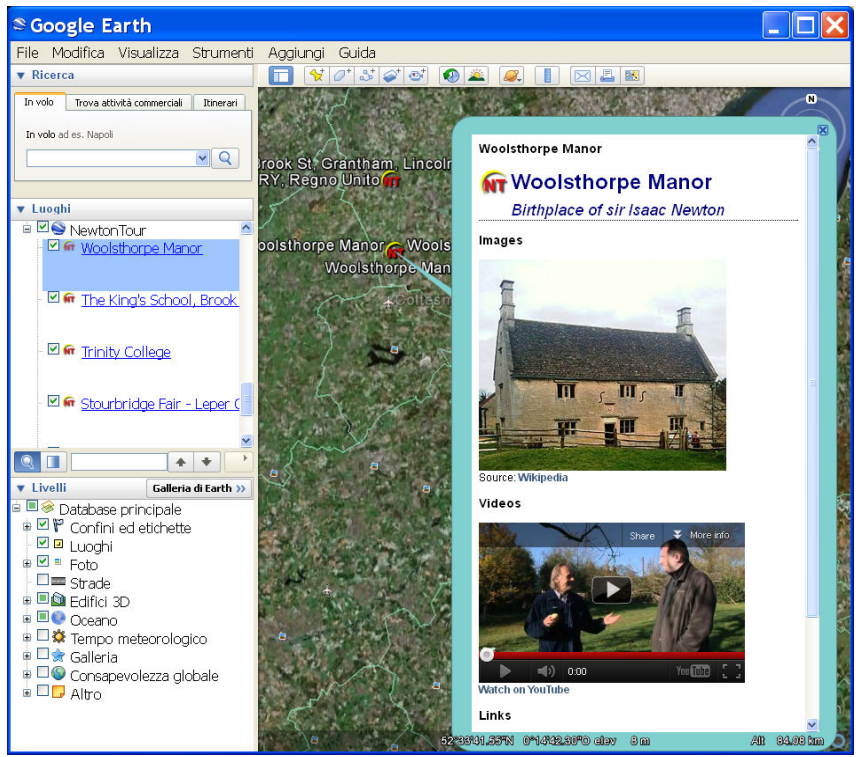 Fig.2

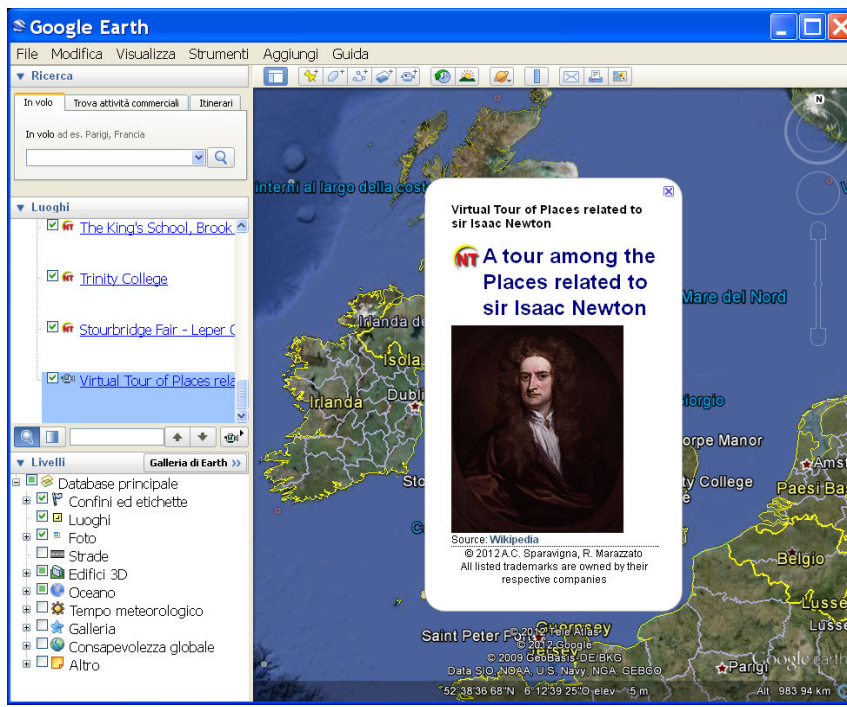 Fig.3

Besides the possibility of inserting static information or motion pictures from YouTube, Google Earth has Street View, representing the possibility to see the places through 360-degree street-level imagery. We have also prepared as movie, labelled as "Virtual Tour of Places related to sir Isaac Newton", which is including the use of Street View. Figure 3 shows what is appearing as soon as the movie starts to run. During the movie, when we visit the Trinity College, we can see some 3D rendering of the place.

In our opinion, the most important feature of KML is the possibility of implementing a comprehensive virtual tour, based on a movie and the use of balloons, which can be run on Google Earth. We proposed these virtual tours to describe the lives of famous persons for teaching purposes. Of course, this method could be implemented to show a real or even to simulate a life, for social purposed. For what concerns Newton's life, a new upgrade of the Newton Tour is under development.